\documentstyle[11pt,epsfig]{article}
\def\edth{\;\raise1.0pt\hbox{$'$}\hskip-6pt\partial\;}
\def\baredth{\;\overline{\raise1.0pt\hbox{$'$}\hskip-6pt
\partial}\;}
\def\gsim{~\rlap{$>$}{\lower 1.0ex\hbox{$\sim$}}}

\def\d{{\rm d}}

\begin{document}

\title{Effect on cosmic microwave background polarization of coupling of
   quintessence to pseudoscalar formed from the electromagnetic
   field and its dual}

\author{Guo-Chin Liu$^1$, Seokcheon Lee$^2$, and Kin-Wang Ng$^{1,2}$}

\maketitle

$^1${\it Institute of Astronomy and Astrophysics,
 Academia Sinica, Taipei, Taiwan 11529, R.O.C.}

$^2${\it Institute of Physics, Academia Sinica,
 Taipei, Taiwan 11529, R.O.C.}

\begin{abstract}
We present the full set of power spectra of cosmic microwave background
(CMB) temperature and polarization anisotropies due to the coupling
between quintessence and pseudoscalar of electromagnetism. This coupling
induces the rotation of the polarization plane of CMB photons, thus
resulting in non-vanishing $B$ mode and parity-violating $TB$ and $EB$ modes.
Using the 2003 flight of BOOMERANG (B03) data, we give the most stringent
constraint on the coupling strength. In some cases, the rotation-induced
$B$ mode can confuse the hunting for the gravitational lensing-induced
$B$ mode.

\end{abstract}

The existence of a dark component with an effective negative
pressure, supported by several observations especially the Hubble
diagram for the type-Ia supernovae (see, for example,
~\cite{Bahcall, Riess}), is still one of the puzzles in cosmology.
Cosmological constant is the simplest possibility for such dark
component. However, the observed value of the cosmological constant
is completely different from theoretical
expectation~\cite{Weinberg}. An alternative candidate, the so-called
quintessence described by a dynamical scalar field $\phi$, is
naturally considered. The dynamics of $\phi$ in general quintessence
models is governed by a scalar potential $V(\phi)$ which makes the
dark energy dominant in recent epoch. There are many different kinds
of proposed potentials, for example, pseudo Nambu-Goldston boson,
inverse power law, exponential, hyperbolic cosine, and tracking
oscillating~\cite{potential}. To differentiate between the models
and finally reconstruct $V(\phi)$ would likely require
next-generation observations.

The quintessential potential $V(\phi)$ and the field $\phi$ itself
are difficult to be measured directly.  What we can do is to
investigate the dark energy density $\Omega_\phi$ and the time
evolution for the equation of state (EOS) $w_\phi=P_\phi/\rho_\phi$,
both of which are governed by the dynamics of $\phi$. Several
observations have been made to give the answers: 157 supernovae in a
redshift interval, $0.015 < z < 1.6$, in the ``Gold Sample''
obtained from a combination of ground-based data and the Hubble
Space Telescope~\cite{Riess} as well as 115 supernovae with $0.015<
z < 1$ from the Supernova Legacy Survey provide information about
the dark energy~\cite{snls}. It is very difficult to determine
whether the quintessence is more preferred than the cosmological
constant by observations of the CMB temperature anisotropy spectrum
only.  Joint analysis of CMB data with supernovae or/and large scale
structure survey such as SDSS or 2dfGRS can offer better
constraints on quintessence models~\cite{joint}. Recently, the study
of the cross-correlation of maps between CMB and various tracers of
matter through the integrated Sachs-Wolfe effect has also been
carried out by several groups~\cite{isw}. However, the intrinsic
properties of dark energy are not well constrained so far from the
investigations above, thus allowing a very wild range of the EOS,
which is strongly model dependent.

An alternative way to study the quintessence is to consider the
interaction of $\phi$ to ordinary matter. Coupling of $\phi$ to dark
matter is considered as a possible solution for the late time
coincidence problem~\cite{coupQ}. It leaves distinct imprints on the
CMB temperature anisotropy and the matter power spectrum due to an
excess of cold dark matter at early epoch
when compared to the standard cosmology model~\cite{Lee}
(We define the "standard cosmology" as the model
without coupling between quintessence and ordinary matter). Of
particular interest in this \emph{Letter}, we study the coupling of
$\phi$ to the pseudoscalar of electromagnetism. Pseudoscalar couplings
usually arise from the spontaneous breaking of a compact symmetry
group, say, U(1) (see Frieman et al. 1995 at
reference~\cite{potential}). Carroll has argued 
that the coupling $\beta_{F\tilde{F}} \phi/\bar{M} F_{\mu \nu}
\tilde{F}^{\mu \nu}$ leads to the rotation of the polarization
vector of propagating photons if $\phi$ is varying with time. Here
$\beta_{F\tilde{F}}$ is the coupling strength, $\bar{M}$ is the
reduced Planck mass, and $\tilde{F}^{\mu \nu}$ is the dual of the
electromagnetic tensor. This effect is called as the ``cosmological
birefringence''~\cite{Carroll1998}.

Measurements of the polarization of distant astronomical objects would
provide information about the cosmological birefringence. Carroll used the
rotation of polarization direction for distant radio sources to
constrain the coupling strength~\cite{Carroll1998}. Another proposed method
is the CMB polarization from the last scattering surface~\cite{Feng1,Feng2}.
In this \emph{Letter}, we study the effects of
the cosmological birefringence on the power
spectra of CMB temperature and polarization. Only assuming the shape
of the potential $V(\phi)$, we present the \emph{first} CMB power
spectra in the presence of the cosmological birefringence by the use
of the full Boltzmann code.

Thomson scatterings of anisotropic radiation by free electrons give
rise to the linear polarization, which is usually described by the
Stokes parameters $Q$ and $U$~\cite{chandrasekar1960}. In standard
cosmology, the time evolution of the polarization perturbation is
governed by the Boltzmann equation~\cite{BEa}. If there is a
physical mechanism which rotates the polarization plane, the
evolution equations for the Fourier modes of the Stokes parameters
are modified to
\begin{eqnarray}
\dot{\Delta}_{Q\pm iU}({\bf k},\eta) &+& ik\mu \Delta_{Q\pm iU}
({\bf k},\eta)=
n_e\sigma_T a(\eta)\Biggl[ -\Delta_{Q\pm iU}({\bf k},\eta)
  \nonumber \\
& & \left .\sum_{m}\sqrt{\frac{6\pi}{5}}
\ _{\pm 2} Y_2^m({\bf \hat n})
S_{P}^{(m)}({\bf k},\eta) \right]
\mp i 2\omega \Delta_{Q\mp iU}({\bf k},\eta),
\label{fps}
\end{eqnarray}
where the derivatives are taken with respect to the conformal time
$\eta$, $\mu={\bf \hat n}\cdot{\bf \hat k}$ is the cosine of the
angle between the CMB photon direction and the Fourier wave vector,
$n_e$ is the number density of free electrons, $\sigma_T$ is the
Thomson cross section, and $a$ is the scale factor. $\ _sY_l^m$ is
spherical harmonics with spin-weight $s$ with $m=0, \pm 1, \pm 2$
corresponding to scalar, vector, and tensor perturbations,
respectively, if the axis of $\ _sY_l^m$ is aligned with the wave
vector ${\bf k}$. $S_P^{(m)}$ is the source term of generating
polarization, which is the composition of the quadrupole components
of the temperature and polarization perturbations $S_P^{(m)}({\bf
k},\eta) \equiv \Delta_{T2}^{(m)}({\bf k},\eta)+
12\sqrt{6}\Delta_{+,2}^{(m)}({\bf k},\eta)+12\sqrt{6}\Delta_{-,2}^
{(m)}({\bf k},\eta)$. We have followed the notation in
Ref.~\cite{Liu} and expanded the perturbations in the spin-0 and
spin-2 spherical harmonics~\cite{NP} according to  the scalar and
tensor properties of the temperature anisotropy and polarization
respectively. Thus, $\Delta_{T2}^{(m)}$ and $\Delta_{\pm,2}^{(m)}$
are the respective expansion coefficients for $\Delta_T$ on basis
$Y_{l,m}$ and $\Delta_{Q\pm iU}$ on basis $\ _{\pm2} Y_{lm}$. The
last term in Eq.~(\ref{fps}) appears due to the rotation of the
polarization plane. The dispersion relation for electromagnetic
radiation coupling to the time varying quintessence field $\phi$ is
given by $E^2=k^2 \pm k\beta_{F\tilde{F}}\dot{\phi}/(a\bar{M})$,
where $\pm$ refer to the right and left handed circular
polarization, respectively. Therefore, the net angular velocity of
the polarization plane is~\cite{Carroll1998}
\begin{equation}
\omega= 2\beta_{F\tilde{F}} \frac{\dot{\phi}}{a\bar{M}}.
\label{angv}
\end{equation}

We are used to decomposing the polarization on the sky into a
divergence free component, the so-called $E$ mode, and a curl
component, the so-called $B$ mode because the values for $Q$ and $U$
depend on the choice of a coordinate system. Whether $B$ mode is
generated depends on the existence of $U$ for the local mode whose
wave vector $ {\bf k}$ parallels to the $\hat{\bf z}$ of the
coordinates whereas $Q$ is defined as the difference in intensity
polarized in the $\hat{\theta}$ and $\hat{\psi}$
directions~\cite{Hu}. Mathematically, for $m=0$, only $Q$ is
generated in the local mode. The axisymmetry of the radiation field
about the mode axis guarantees that no $B$ mode can be generated by
scalar mode perturbations.

In the presence of cosmological birefringence, we can find two
important features in Eq.~(\ref{fps}). Firstly, the rotation of the
polarization plane generates $U$ contributions to the local mode
polarization. This converts the power from the $E$ mode to $B$ mode.
The conversion depends on how much rotation there is from the epoch
when the polarization is generated to today. Secondly, $TB$ and $EB$
cross correlations are expected to vanish due to the parity ($T$ and
$E$ have parity $(-1)^\ell$ while $B$ has $(-1)^{\ell+1}$). The
cosmological birefringence violates the parity and thus generates
the $TB$ and $EB$ power spectra whose magnitudes depend on the
integrated rotation of polarization too. Substituting the angular
velocity of the polarization plane in Eq.~(\ref{angv}) into
Eq.~(\ref{fps}), we calculate the power spectra $T$, $E$, $B$, $TE$,
$TB$, and $EB$ modes.

These six power spectra form a complete two-point statistics of CMB
temperature and polarization anisotropies. To simplify the calculation,
we only include the scalar perturbations.  That is by setting $m=0$ in
Eq.~(\ref{fps}). Without showing the details, we just write down
the power spectra which are obtained from the solutions for the
line-of-sight integration as
\begin{eqnarray}
C^{(E,B)}_{\ell}&=&(4\pi)^2\frac{9}{16}\frac{(\ell+2)!}{(\ell-2)!} \int
k^2 \d k \left [ \Delta_{(E,B)}(k, \eta_0)\right ]^2 , \nonumber \\
C^{EB}_{\ell}&=&(4\pi)^2\frac{9}{16}\frac{(\ell+2)!}{(\ell-2)!} \int
k^2 \d k   \Delta_{E}(k, \eta_0) \Delta_{B}(k, \eta_0)  , \nonumber \\
C^{TE}_{\ell}&=&(4\pi)^2\sqrt{\frac{9}{16}\frac{(\ell+2)!}{(\ell-2)!}} \int
k^2 \d k  \Delta_T (k, \eta_0) \Delta_E (k, \eta_0) , \nonumber \\
C^{TB}_{\ell}&=&(4\pi)^2\sqrt{\frac{9}{16}\frac{(\ell+2)!}{(\ell-2)!}} \int
k^2 \d k  \Delta_T (k, \eta_0) \Delta_B (k, \eta_0),
\label{ps}
\end{eqnarray}
where
\begin{eqnarray}
\Delta_T (k, \eta_0)&=&\int_0^{\eta_0} \d\eta g(\eta) S_T(k,\eta){j_\ell(kr)},
\nonumber \\
\Delta_{E}(k, \eta_0)+i\Delta_{B}(k,\eta_0) &=& \int_0^{\eta_0}
\d\eta g(\eta) S_P(k,\eta) \frac{j_\ell(kr)}{(kr)^2}
e^{i2\alpha(\eta)},
\end{eqnarray}
where the visibility function $g(\eta)$ describes the probability
that a photon scattered at epoch $\eta$ reaches the observer at the
present time, $\eta_0$. Similar to $S_P\equiv S_P^{(0)}$,
$S_T$ is the source term generating the temperature
anisotropy. $j_\ell$ is the spherical Bessel function and
$r=\eta_0-\eta$. The rotation angle $\alpha(\eta)=\int_\eta^{\eta0}
d\eta' \omega(\eta')$. We do not present the formula for the temperature
anisotropy because it is unchanged under the rotation of the
polarization plane.

We have modified the public code CMBFast~\cite{SZ} for our purpose.
Here, we consider the potential $V(\phi)=V_0 \exp(\lambda \phi^2
/2\bar{M}^2)$ for our quintessence model (the hyperbolic cosine
potential is also considered, see below), where $\lambda$ is a
parameter determining how shallow the potential is. Hereafter we fix
$\lambda = 5$ and we will obtain similar results by choosing other
values of $\lambda$. We plug the table of the EOS for this
quintessence model into the modified CMBFast code and input the
cosmological parameters from the best fit values of the WMAP
three-year results~\cite{spergel}. The power spectra are then
normalized to the first peak of the temperature anisotropy measured
by the three-year WMAP observation~\cite{Hinshaw}. The left panel of
Fig.~1 shows the $E$ and $B$ mode power spectra with the coupling
strength $\beta_{F\tilde{F}}$ ranging from $10^{-5}$ to $10^{-3}$.
The $EB$ mode power spectrum is shown in the right panel. On small
scales, increasing coupling strength results in a suppression of the
$E$ mode in the standard model and non-vanishing $B$ and $EB$ modes.
Furthermore, the shapes of the $B$ and $EB$ mode power spectra
basically follow the standard $E$ mode except the reionization bump
on large scales. To explain this, we may make a rough estimation in
Eq.~(\ref{ps}): $C_{B\ell}\sim C_{E\ell}\sin^2 2\alpha_\ell$ and
$C_{EB\ell}\sim 0.5C_{E\ell}\sin 4\alpha_\ell$ with $\alpha_\ell$
being the total rotated angle for certain angular scales $\theta
\sim \pi/\ell$ from last scattering epoch to today. This
$\alpha_\ell$, in general, is not constant for all the scales. The
$E$ mode power on small scales mainly comes from the recombination
epoch at $z \sim 1100$. On the other hand, the boosting power on
large scales comes from reionization epoch when the CMB photons are
rescattered by free electrons at $z \sim 10$~\cite{page}. From
Eq.~(\ref{angv}) and the evolution of $\phi$, we find that the
integrated rotation angle from the reionization epoch is much
smaller than that from the recombination epoch. Therefore, there is
much less power converted from $E$ mode to $B$ mode on large scales
than small scales.

\begin{figure}[htbp]
\centerline{\psfig{file=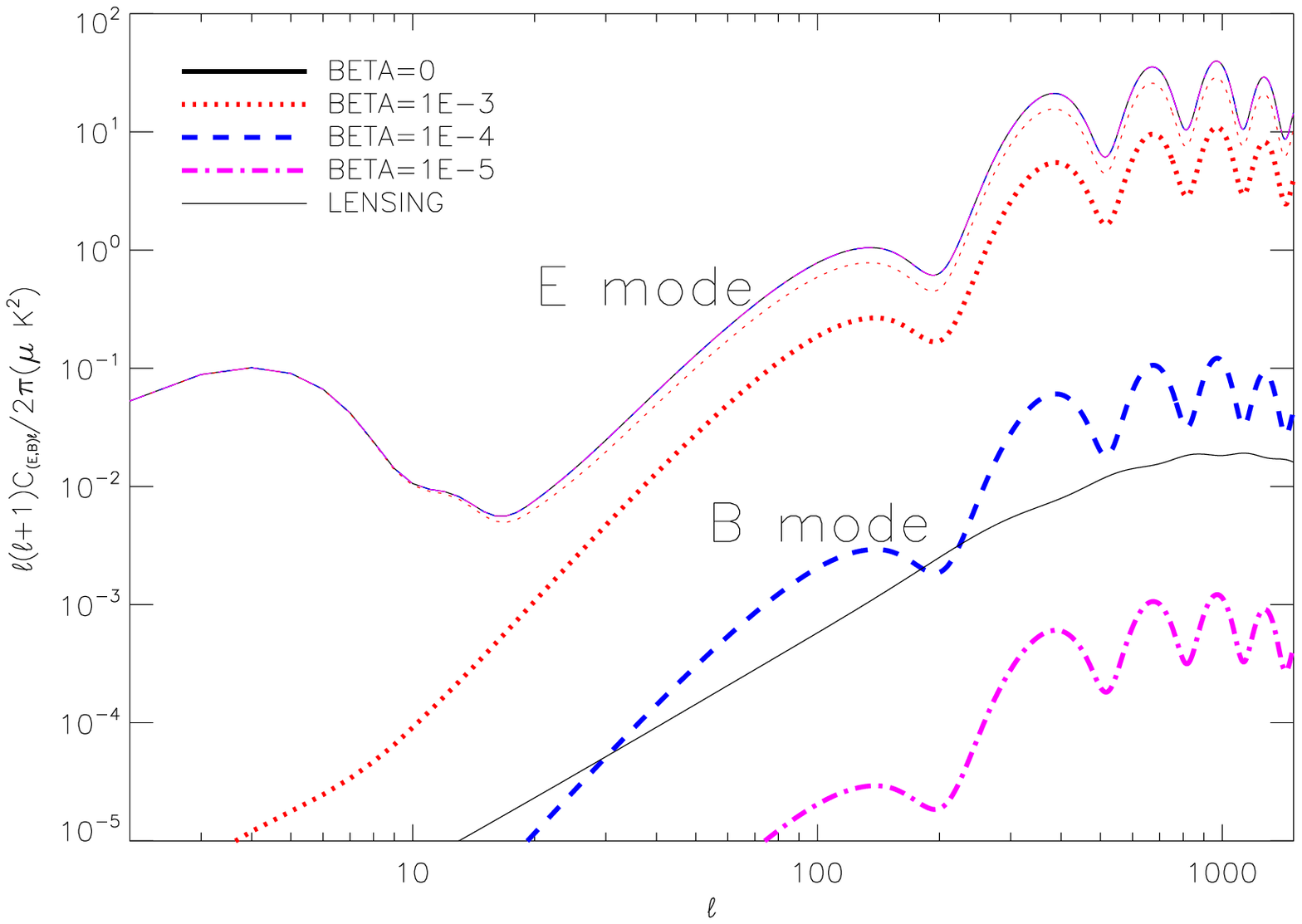, width=5.5cm}
\psfig{file=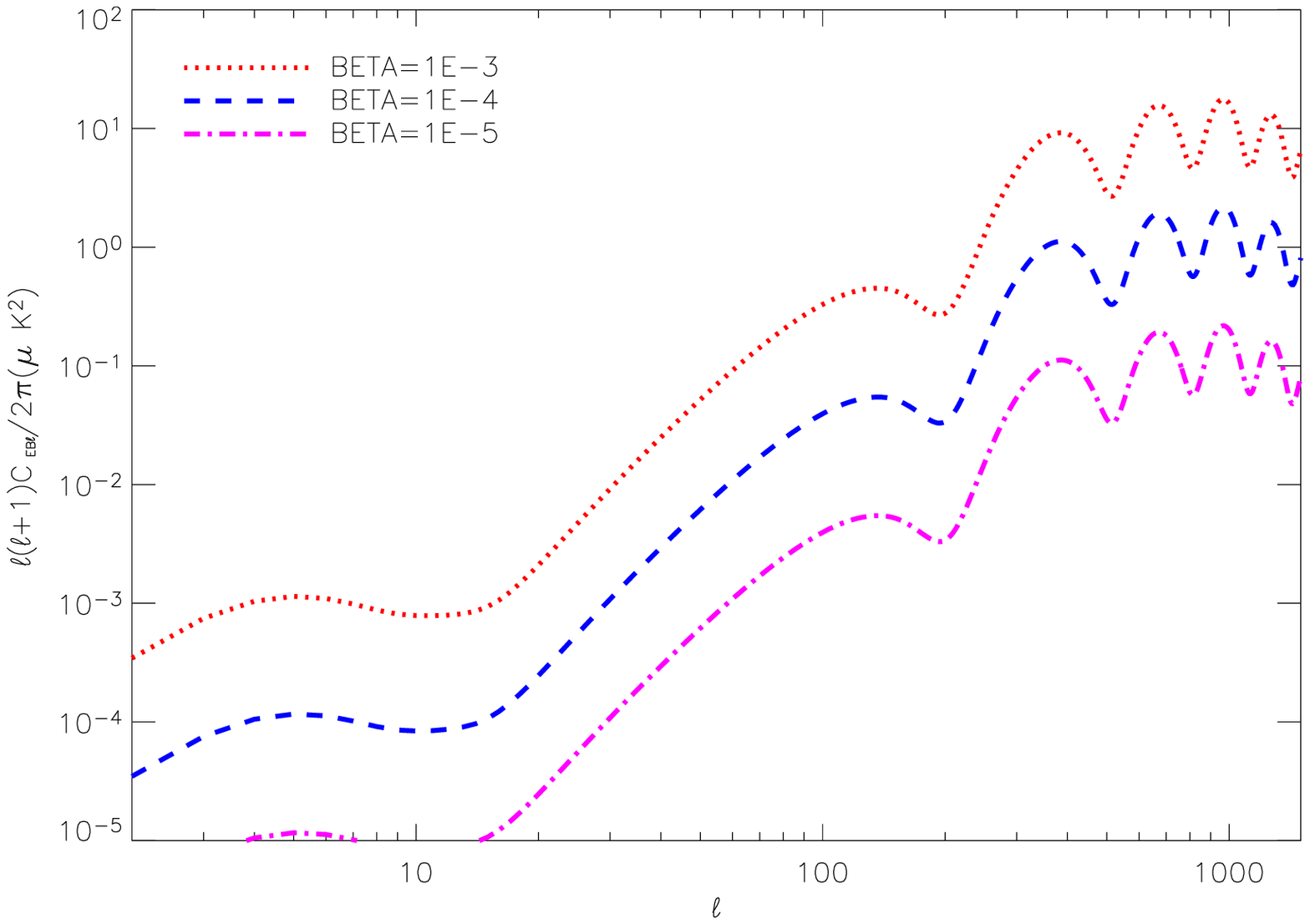, width=5.5cm}} \label{fig:CBL} \caption{$E$,
$B$ (left panel; the lower three thick curves are $B$ modes) and
$EB$ (right panel) mode power spectra from the cosmological
birefringence with different coupling strength.}
\end{figure}

If the coupling strength is large enough, the $B$ mode induced by
the cosmological birefringence will mix up with the gravitational
lensing-induced $B$ mode. Gravitational lensing by large scale
structures modifies slightly the primary $E$ mode power spectrum.
Most noticeably it generates, through mode coupling, $B$ mode
polarization out of pure $E$ mode signal~\cite{Benabed}. The
lensing-induced $B$ power spectrum, which peaks around $\ell \sim
1000$, has the roughly similar shape with that from the
birefringence. We also show the power spectrum of the
lensing-induced $B$ mode in Fig.~1 by a thin solid curve for
comparison. The birefringence-induced $B$ mode is indeed compatible
with the lensing-induced $B$ mode for $\beta_{F\tilde{F}} \sim
10^{-4}$.

\begin{figure}[htbp]
\centerline{\psfig{file=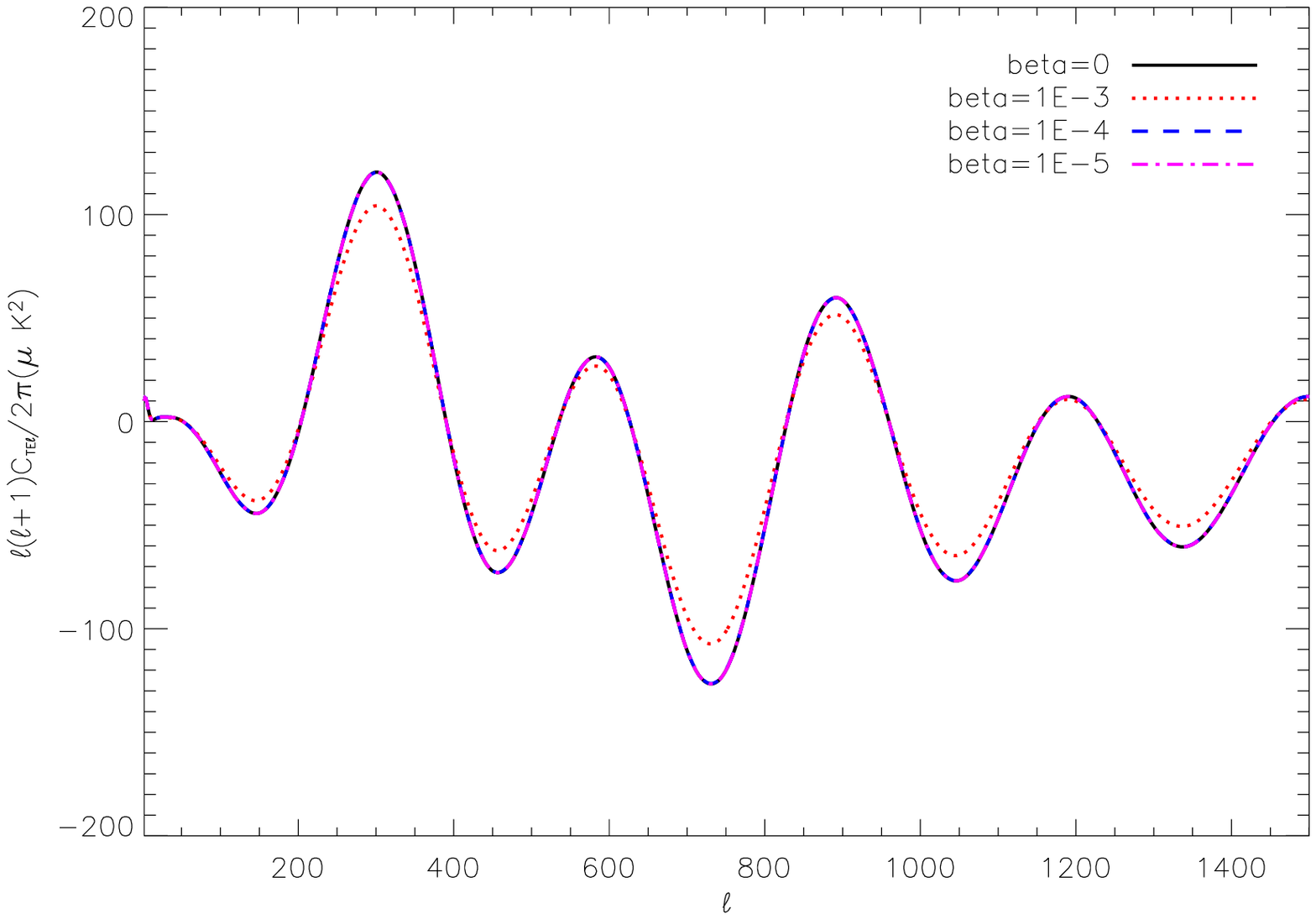, width=5.5cm}
\psfig{file=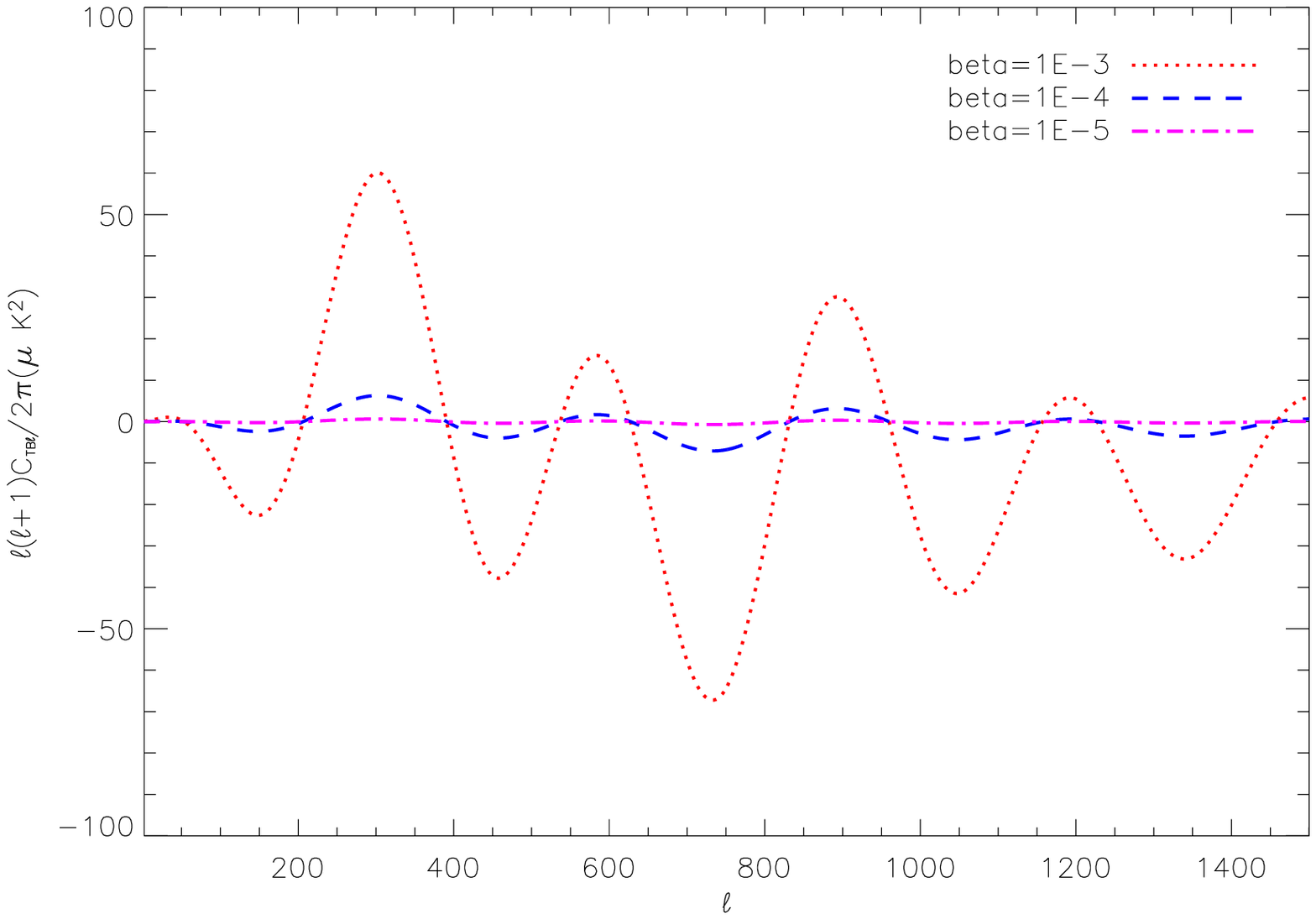, width=5.5cm}}\label{figCTBL} \caption{$TE$
(left panel) and $TB$ (right panel) mode power spectra from the
cosmological birefringence with different coupling strength.}
\end{figure}

Fig.~2 shows the $TE$ and $TB$ power spectra for different coupling
strength. Having the complete set of power spectra, we can constrain
the coupling strength from observed data. In order to focus on the
cosmological birefringence, we do not make the global fit to all the
cosmological parameters. It is debatable that $\beta_{F\tilde{F}}$
is degenerate with other cosmological parameters due to the
decrement of the $TE$ and $E$ power spectra. As we will see later,
the upper limit on the coupling strength prevents it from making a
significant change on the $TE$ and $E$ modes.

\begin{figure}[htbp]
\centerline{\psfig{file=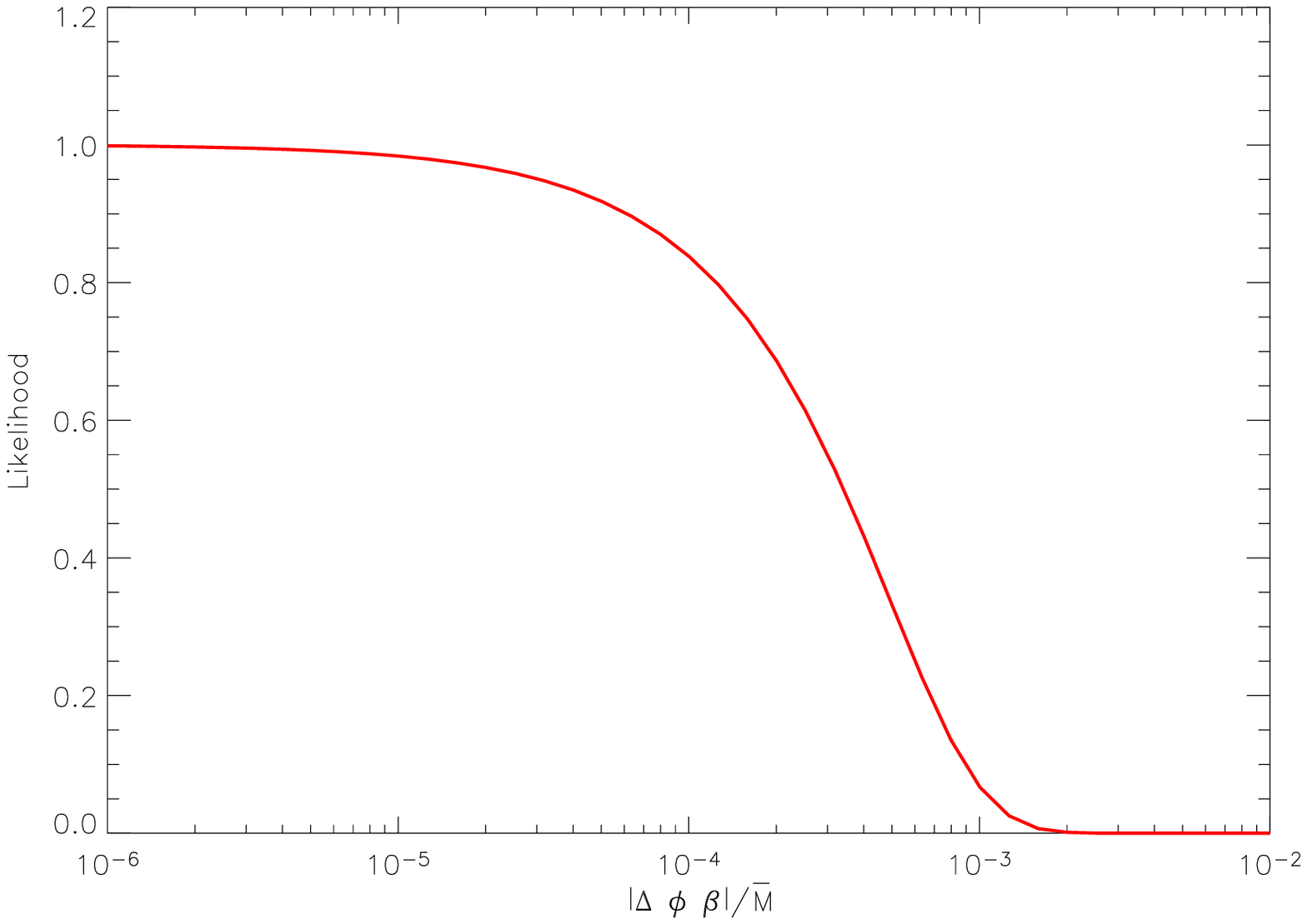, width=5.5cm}}
\label{likeli}\caption{Likelihood function of the coupling
strength.}
\end{figure}

We use the North American data from the 2003 Antarctic flight of
BOOMERANG~\cite{boom} to calculate the $\chi^2=\sum_{i,j}
(D_i^b-T_i^b) C^{-1}_{i,j} (D_j^b-T_j^b)$ fitting for
$\beta_{F\tilde{F}}$ while fixing other cosmological parameters,
where $D_i^b$ is the $i$th band-power, $C$ is the covariance matrix,
and $T_i^b=\sum_\ell C_{\ell} W_{i\ell}/\ell$ is obtained by the
theoretical prediction multiplied by the band-power window function
$ W_{i\ell}/\ell$. Those band-power data, including covariance
matrices and window functions, are available; see~\cite{boom_data}. In
practice, we only fit 
$\beta_{F\tilde{F}}$ to the $EB$ and $TB$ power spectra which come
from the parity violation. We do not use the $B$ mode power spectrum
because it is contaminated by lensing-induced signal. If the
lensing-induced $B$ mode can be successfully cleaned by appropriate
techniques such as those proposed by Seljak and Hirata~\cite{SH}, we
expect that including the $B$ mode will give a stronger constraint.
We convert $\chi^2$ to the likelihood by ${\cal L}=e^{-\chi^2/2}$
and normalize the maximum likelihood value to unity. The result is
shown in Fig.~3. The upper limit on the coupling strength is found
to be $|\beta_{F\tilde{F}} \Delta \phi|/\bar{M} < 8.32 \times
10^{-4} $ at $95\%$ confidence level, where $\Delta \phi$ is the
total change of $\phi$ till today. This small value of the coupling
strength gives an insignificant change on $TE$ and $E$ modes and
thus will not affect the determination of the cosmological
parameters. We also use the hyperbolic cosine potential $V(\phi)=V_0
\cosh(\lambda \phi/\bar{M})$ for the testing. Even though the
quintessence evolution in this potential is different from that in
the exponential case, the upper limit value of $|\beta_{F\tilde{F}}
\Delta \phi|/\bar{M}$ does not change much.
%We have also test other potentials. Even though the dynamics
%evolution of $\phi$ are different in each case, the upper limit dose
%not change much for different cases.
This value is much smaller than the result in
Ref.~\cite{Carroll1998}, $3\times 10^{-2}$, where $\Delta \phi$ is
only from $z=0.425$ to today. It is remarkable that the
rotation-induced $B$ mode with the upper limit value of the
coupling strength exceeds the lensing-induced $B$ mode.
Therefore, careful measurements of $TB$ and $EB$ are necessary
for separating the two effects.

Several authors
have studied the effect of parity violation on the CMB power spectra
by assuming a constant rotation angle $\alpha$~\cite{Feng1,Feng2}.
They obtained a new set of rotated power spectra from combining the
power spectra in the standard model with the sine or cosine function
of $\alpha$. Furthermore, taking the rotation angle as a free
parameter, Feng et al.~\cite{Feng2} constrained the rotation angle
using the data made by the first year WMAP and BOOMERANG
observations. However, the time-varying scalar field $\phi$ in their
work is constrained such that the integrated rotation angle should
be very small from the recombination to the reionization epoch.
Therefore, it is less supported by general quintessence models.

We are grateful to BOOMERANG team for the use of data, especially to
F. Piacentini for making $TB$ and $EB$ covariance
matrices~\cite{boom}. We acknowledge anonymous referees for the useful
comments to improve the \emph{Letter}. G.C.L. thanks K. Ichiki for
fruitful discussion. K.W.N. was supported in part by NSC grant
NSC94-2112-M-001-024.

\end{document}